\documentclass[aps,pra,twocolumn,amsmath,amssymb,showpacs,superscriptaddress]{revtex4-1}
\usepackage{color}
\usepackage{textcomp} 
\usepackage{mathrsfs,amsmath}
\usepackage{graphicx}
\usepackage{dcolumn}
\usepackage[mathlines]{lineno}
\usepackage{hyperref}
\usepackage{bm}
\usepackage{epstopdf}
\usepackage{soul}
\usepackage{epsfig}
\usepackage{bbold}

\usepackage{booktabs}

\usepackage{changes}
\usepackage{lipsum}
\setremarkmarkup{(#2)}

\usepackage{color}

\begin{document}

\title{Enhancement of the second-harmonic generation based on the cascaded
	second- and third-order nonlinear processes in a multimode optical
	microcavity}

\author{Ming Li}
\affiliation{Centre for Optical and Electromagnetic Research, JORCEP, State
	Key Laboratory for Modern Optical Instrumentation, Zhejiang Provincial
	Key Laboratory for Sensing Technologies, Zhejiang University, Hangzhou
	310058, China}
\author{Chang-Ling Zou}
\email{clzou321@ustc.edu.cn}
\affiliation{Key Laboratory of Quantum Information, CAS, University of
	Science and Technology of China, Hefei 230026, China}
\author{Chun-Hua Dong}
\affiliation{Key Laboratory of Quantum Information, CAS, University of
	Science and Technology of China, Hefei 230026, China}
\author{ Xi-Feng Ren}
\affiliation{Key Laboratory of Quantum Information, CAS, University of
	Science and Technology of China, Hefei 230026, China}
\author{Dao-Xin Dai}
\email{dxdai@zju.edu.cn}
\affiliation{Centre for Optical and Electromagnetic Research, JORCEP, State
	Key Laboratory for Modern Optical Instrumentation, Zhejiang Provincial
	Key Laboratory for Sensing Technologies, Zhejiang University, Hangzhou
	310058, China}

\begin{abstract}
	Optical microcavities are often used to realize enhanced nonlinear
	optical interactions for highly efficient second-harmonic generation.
	With increased pump power, the efficiency of nonlinear frequency conversion
	can be increased further, while some other unwanted nonlinear effects
	will also emerge, leading to complicated dynamics or instability.
	Here, we study the interplay between cascaded second- and third-order
	nonlinear processes and investigate their impact on the second-harmonic
	generation in microcavities. It is found that the non-degenerate optical
	parametric oscillation (OPO) appears and the presence of $\chi^{(3)}$
	process can modify the OPO threshold significantly when the multimode
	cavity is strongly pumped at the fundamental optical mode. One can
	even break the efficiency limitation of the second-harmonic mode restricted
	by the OPO by utilizing the interference between the OPO and the four-wave
	mixing. The present coherent interplay between nonlinear optical processes
	in microcavities is conducive to exploring new physics in the cavity
	nonlinear photonics.
\end{abstract}
\maketitle

\section{Introduction}

Cavity-assisted nonlinear optical effects have been studied since
the last century for their applications in parametric amplification
\cite{giordmaine1965tunable}, frequency conversion \cite{pereira1988generation,ou199285}
and frequency combs \cite{bartels200910}, by putting nonlinear crystals
in traditional macroscopic Fabry-Perot or ring-type cavities. Along
with the development of fabrication technologies for high-$Q$ microcavities
\cite{akahane2003high,vahala2003optical,matsko2006optical}, strong
nonlinear optical effects have been demonstrated \cite{vahala2003optical,strekalov2016nonlinear,Lin:17},
in which the nonlinear interaction between photons of different colo	rs
is greatly enhanced due to the small mode volume of microcavities.
Enhanced nonlinear photonics in microcavities enables low-threshold
lasers \cite{min2003compact,latawiec2015chip}, optical parametric
oscillators \cite{saba2001high}, optical squeezing\cite{dutt2015chip,dutt2016tunable},
frequency combs \cite{del2007optical,kippenberg2011microresonator,jung2014green,herr2014temporal,huang2015mode,chembo2016quantum},
spontaneous down-conversion quantum photon sources \cite{clemmen2009continuous,guo2014telecom,grassani2015micrometer,silverstone2015qubit,guo2017parametric},
highly efficient frequency conversion \cite{Guo2016a,li2016efficient,rueda2016efficient}
as well as second-harmonic and third-harmonic generation \cite{carmon2007visible,yang2007enhanced,rivoire2009second,levy2011harmonic,pernice2012second,kuo2014second,Guo2016,lin2016cavity}. 

As has been demonstrated in the previous researches, different nonlinear
processes might happen simultaneously in an optical microcavity \cite{Lin:17,moore2011continuous,wolf2017cascaded}.
For example, parametric down-conversion can be observed along with
the second-harmonic generation (SHG) \cite{liu2017cascading}. This
type of cascaded second-order ($\chi^{(2)}$) nonlinear processes
have been studied a lot in both classical and quantum fields \cite{Marte1994,white1997classical}.
By harnessing the cascaded effect of the $\chi^{(2)}$ nonlinearity,
it is feasible to achieve efficient higher-order harmonic light generation
\cite{Sasagawa2009,moore2011continuous,levy2011harmonic,liu2017cascading},
multipartite entanglement \cite{Tan2011} and frequency combs in
different mode families \cite{wu2012optical,ulvila2013frequency,ricciardi2015frequency}.
However, for the analysis of those nonlinear optical processes, previously
people usually use a simple model, in which only the $\chi^{(2)}$
nonlinearity is taken into consideration. Such a simple model is only
valid when the other nonlinear interactions such as $\chi^{(3)}$
and Raman are very weak, or the interactions involving other modes
are not on-resonance \cite{Guo2016}. In presence of the $\chi^{(3)}$,
Raman or other types of nonlinear processes, the $\chi^{(2)}$ nonlinear
process may be affected. As a result, the simple model does not work
well .

For example, for the SHG, the high SH power will dissipate to other
modes through spontaneous down-conversion, and the pump power may
also dissipated to other modes through the spontaneous four-wave mixing
(FWM). As a result, the high conversion efficiency predicted by the
simple model may be invalid. A fundamental question for the multimode
cavity nonlinear photonic system is: what is the efficiency of SHG
in a cavity supporting both $\chi^{(2)}$ and $\chi^{(3)}$ nonlinearity
and will the different nonlinear processes interfere with each other?
In this work, we establish an improved model containing both the $\chi^{(2)}$
and $\chi^{(3)}$ nonlinear interactions in a multimode microcavity.
With this model, we investigate the interference between cascaded
second-order process and third-order process and its
effect on SHG. According to our calculations, the $\chi^{(3)}$ process
can modify the threshold of parametric down-conversion, thus altering
the energy flow from the second-harmonic (SH) mode to its neighbor
modes. It is revealed that a key for achieving higher SHG efficiency
is to optimize the resonant frequency of the SH mode according to
the coupling strength of the $\chi^{(3)}$ process. 

\section{Interference between $\chi^{(2)}$ and $\chi^{(3)}$ processes}

Fig.$\,$1(a) shows the considered microcavity system, where multiple
optical modes in both the visible and infrared bands are supported
in the same cavity. When the cavity is designed carefully for phase
matching and tuned precisely for double resonances, efficient SHG
can be achieved, as demonstrated experimentally in \cite{Guo2016}.
The photon interaction Hamiltonian is described by a $\chi^{(2)}$
process between mode $a$ and $d$ as $a^{2}d^{\dagger}+a^{\dagger2}d$.
Traditionally, most work in cavity-enhanced SHG was based on the double-resonance
condition and the non-depletion approximation \cite{yang2007enhanced,rivoire2009second,pernice2012second}. For the
the case with weak pump power and weak nonlinear effect, the spontaneous
parametric down-conversion (SPDC) and FWM processes induced by the
vacuum fluctuation can be neglected. Therefore, the SHG process is
often approximately modeled as two coupled harmonic oscillators. 

\begin{figure}
	\begin{centering}
		\includegraphics[width=8cm]{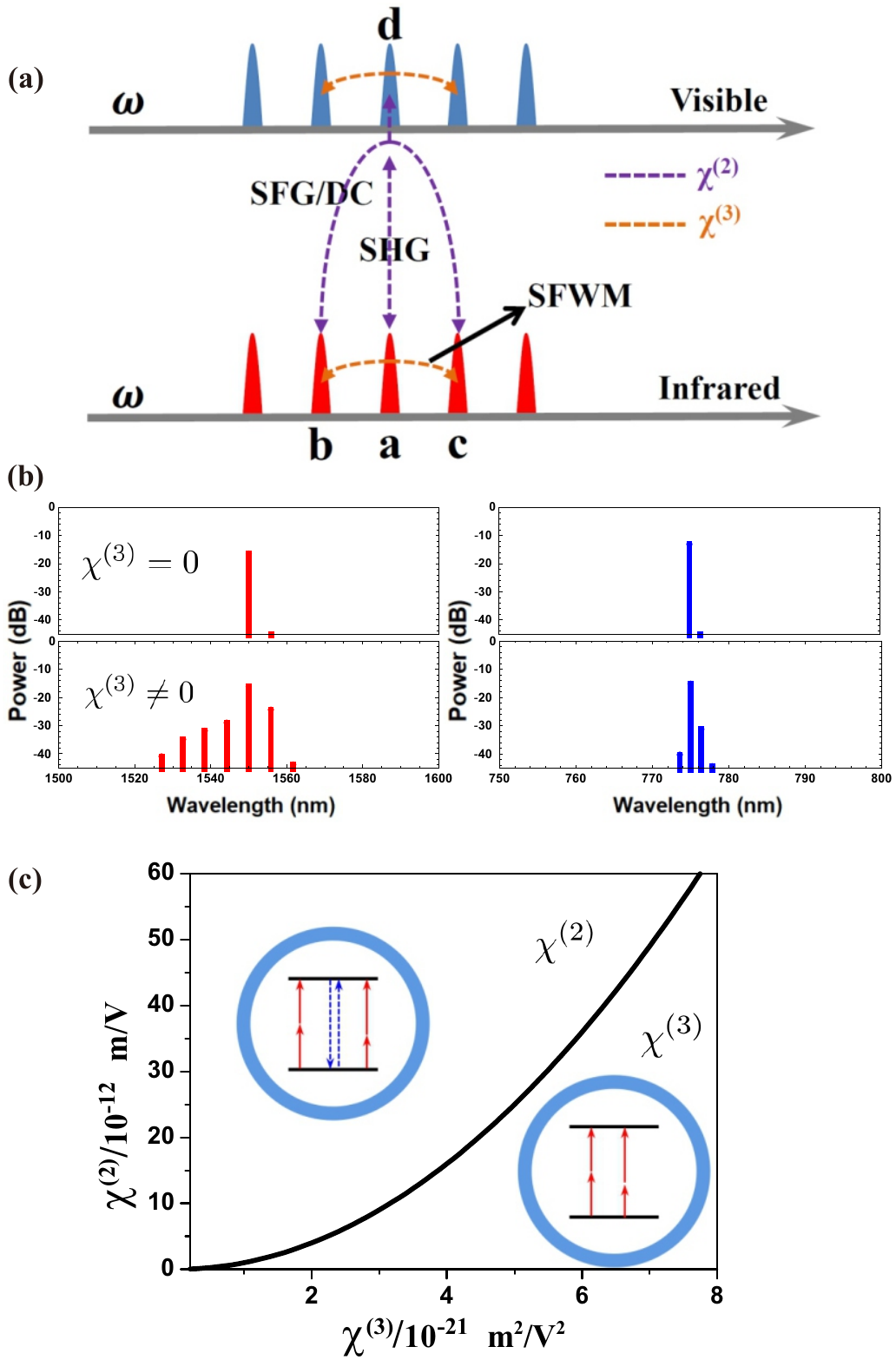}
		\par\end{centering}
	\caption{(a) Schematic picture of hybrid nonlinear processes in a cavity. FWM:
		Four-wave mixing between $a$ and $b,c$. SHG: Second-harmonic generation
		between $a$ and $d$. SFG/DC: Sum-frequency generation/down conversion
		between $d$ and $b,c$. (b) Simulated energy distribution
		in two mode families. Both cases are obtained by scanning the detuning
		of the two mode families. For the case $\chi^{(3)}=0$, the system
		stays below the OPO threshold for all detunings. While $\chi^{(3)}$
		nonlinearity pushes the system above threshold. (c)  Comparison between the coupling strengths of cascaded $\chi^{(2)}$-$\chi^{(2)}$ and $\chi^{(3)}$ processes.
		Here we set $\frac{3}{8\epsilon}\frac{\xi_{2}^{2}}{\xi_{3}}=2.5\times10^{-2}$,
		$\omega_{1}/2\pi=2\times10^{14}\:Hz$. The two processes will interfere
		with each other around the black line with $Q=10^{5}$. For higher
		quality factors, the $\chi^{(2)}$ process dominates. The insets are
		the energy diagram of cascaded $\chi^{(2)}$-$\chi^{(2)}$ process
		(up) and $\chi^{(3)}$ process (down). }
\end{figure}

However, one should notice that there also exist multiple adjacent
modes near mode $a$ and $d$ in a cavity. The orbit angular momentum
of these modes differ by integers and their frequencies are almost
equally spaced. When the pump power increases, the intra-cavity photon
number arrives at the threshold of stimulated emission at mode $a$
due to the degenerate parametric down conversion process $a^{2}d^{\dagger}+a^{\dagger2}d$
(a reversal process of SHG), breaking the non-depletion condition
of SHG. Also, $d$ can couple to the adjacent modes $(b,c)$ around
the fundamental mode $a$ (see Fig.$\:$1(a)) through non-degenerate
SPDC ($bcd^{\dagger}+b^{\dagger}c^{\dagger}d$). Both processes might
decrease the efficiency of SHG. 

In addition to the cascaded SHG and down-conversion, the $\chi^{(3)}$
process redistributes the energy in the modes adjacent to $a$ and
$d$ through the FWM. Taking the FWM, self-phase modulation (SPM) and cross-phase modulation (XPM) into consideration, the system can be described by the Hamiltonian \cite{xiangguo},
\begin{eqnarray}
H & = &\sum_{0}^{N_{1}}\omega_{f,j}f_{j}^{\dagger}f_{j}+\sum_{0}^{N_{2}}\omega_{s,j}s_{j}^{\dagger}s_{j}\nonumber\\
&  & H_{\chi^{(2)}}+H_{\chi^{(3)}}\label{eq:Ham_full},
\end{eqnarray}
where $f_{j}$ is the $j$-th mode in the fundamental mode familiy, $s_{j}$ is the $j$-th mode in the SH mode familiy, $H_{\chi^{(2)}}=\sum_{j,k,l}g_{2}^{jkl}(f_{j}f_{k}s_{l}^{\dagger}+f_{j}^{\dagger}f_{k}^{\dagger}s_{l})$ is the interaction due to $\chi^{2}$,  $H_{\chi^{(3)}}$ is the interaction due to $\chi^{3}$ including the four-wave mixing in one mode familiy and between two mode families, which has the form $g_{3}^{jklm}(f_{j}f_{k}s_{l}^{\dagger}s_{m}^{\dagger}+s_{j}s_{k}f_{l}^{\dagger}f_{m}^{\dagger})$. $\omega_{f,j}$ is the resonant frequency of the $j$-th mode in the fundamental mode family and similar for the SH mode, $N_{1}$ and $N_{2}$ are the number of modes in each mode familiy, $g_{3}$ and $g_{2}$ represent the single photon coupling strength of the third- and second-order process.

In order to quantitatively analyze the system, we estimate the single
photon coupling strengths of different processes. In a ring-type microcavity,
the single photon coupling strength can be derived from the nonlinear
interaction Hamiltonian $E\cdot P_{nl}$, which are given in the following
for SHG ($\chi^{(2)}$) and degenerate FWM ($\chi^{(3)}$) \cite{strekalov2016nonlinear,Guo2016},
\begin{eqnarray}
\text{\ensuremath{\hbar}}g_{2} & = & \sqrt{\frac{\hbar\omega_{1}\hbar\omega_{1}\hbar\omega_{2}}{\epsilon_{0}\epsilon_{1}^{2}\epsilon_{2}}}\frac{3\chi^{(2)}}{4\sqrt{2}}\frac{1}{\sqrt{2\pi R}}\frac{1}{\sqrt{A_{eff}}}\xi_{2}\label{eq:g2}
\end{eqnarray}
\begin{eqnarray}
\hbar g_{3} & = & \sqrt{\frac{(\hbar\omega_{1})^{2}\hbar\omega_{2}\hbar\omega_{3}}{(\epsilon_{0}\epsilon_{2})^{2}\epsilon_{1}\epsilon_{3}}}\chi^{(3)}\frac{3}{2}\frac{1}{2\pi R}\frac{1}{A_{eff}}\xi_{3}\label{eq:g3}
\end{eqnarray}
where $R$ is the radius of the microcavity, $\xi_{i}$ is the mode overlap factor in the cross section and fulfills
$0\leq\xi_{i}\leq1$, $A_{eff}$ is the effective mode area in the
cross section, $\epsilon_{i}$ is the relative dielectric constant
at frequency $\omega_{i}$.

To give a preliminary proof of the complicated multimode and multiple
nonlinearity processes, we numerically simulate the coupling between
the $1550\:nm$ fundamental frequency mode family ($30$ modes) and
$775\:nm$ SH mode family ($30$ modes) in a AlN microcavity based the Hamiltonian in Equation (\ref{eq:Ham_full}). The parameters used in the simulation are $g_{2}=0.1\:MHz$, $g_{3}=5\:Hz$, $D_{1}^{1550}=4.57\times10^{12}\:Hz$, $D_{2}^{1550}=1.76\times10^{9}\:Hz$, $D_{1}^{775}=4.42\times10^{12}\:Hz$, $D_{2}^{775}=2.06\times10^{9}\:Hz$, where the dipsersion parameter $D_{1}$ and $D_{2}$ are derived from the Taylor expansion of the resonance frequency $\omega_{j}=\omega_{0}+jD_{1}+\frac{1}{2!}D_{2}+...$. These parameters are referred to the experimental work in Ref.\cite{xiangguo}. As shown
in Fig.$\:$1(b), without the $\chi^{(3)}$ interaction, there are
only two modes generated and the system operates below the OPO threshold
(the critical point that the adjacent modes just start to oscillate).
When there exists the $\chi^{(3)}$ interaction, OPO appears and many
modes are excited in each mode family. As a result, the energy flows
to mode $b,c$ and their adjacencies, and the SHG efficiency decreases.
The preliminary analysis shows that the $\chi^{(2)}$ process is indeed
influenced by the $\chi^{(3)}$ process.

Considering that the visible modes have lower
$Q$ and the pump is in the infrared band, it is reasonable to neglect
the FWM in the visible band. Due to the dispersion of the cavity mode, resonant modes far from the pump have relatively large detunings. This is also verified by the results in Fig.$\:$1(b) that only a few of modes near $d$ in the visible band is significantly involved in the
nonlinear processes.  Therefore, there are mainly three processes in the microcavity: 

(i) $\chi^{(2)}$ process $\left(a+a\rightarrow d\right)$, with a
coupling strength $g_{2}$.

(ii) $\chi^{(3)}$ process $\left(a+a\to b+c\right)$ with a coupling
strength $g_{3}$.

(iii) cascaded $\chi^{(2)}$-$\chi^{(2)}$ process ($a+a\rightarrow d\rightarrow b+c$) \cite{Marte1994}.
The effective coupling strength scales with $g_{2}^{2}/\left(-i\delta_{d}-\kappa_{d}\right)$,
where $\delta_{d}$ and $\kappa_{d}$ are the detuning and decay rate
of mode $d$, respectively. 

Generally, the $\chi^{(3)}$ susceptibility of non-centrosymmetric
materials is $8$ to $10$ order of magnitude smaller than $\chi^{(2)}$
\cite{boyd2003nonlinear}. The third-order nonlinear interaction
of optical modes is negligible compared to second-order interaction
for most situations in traditional bulk, fiber and waveguide optics.
Things may change when we study the nonlinear interaction in an optical
cavity. The highly concentrated optical field in both space and time
domain will significantly enhance the interaction strengths in different
manners for different processes. The processes (ii) and (iii) are
negligible for small pump power, and should be take into consideration
when the effective cooperativity $C^{(ii)}=g_{3}^{2}N_{a}^{2}/\kappa_{b}\kappa_{c}$
and $C^{(iii)}=g_{2}^{2}N_{d}/\kappa_{b}\kappa_{c}$ approach unity,
where $N_{d}\lesssim4g_{2}^{2}N_{a}^{2}/\kappa_{d}^{2}$, $\kappa_{i}$
is the decay rate of the $i$-th mode with $i\in\{a,b,c,d\}$. Under
certain conditions, the probability amplitudes of these processes
might have the same order of magnitude. The phase between the two
paths will cause constructive or destructive interference. 

When the coupling strength $g_{3}$ and $g_{2}^{2}/\kappa_{d}$ are
comparable, i.e. $C^{(ii)}\sim C^{(iii)}$, the interference of the
three nonlinear processes occurs if the pump (or intracavity photon
number $N_{a}$ in the pump mode) is large enough. This condition
requires 
\begin{equation}
\kappa_{d}\simeq\frac{g_{2}^{2}}{g_{3}}\simeq\frac{3\omega_{1}}{8\epsilon}\frac{\xi_{2}^{2}}{\xi_{3}}\frac{\chi^{(2)}{}^{2}}{\chi^{(3)}}.
\end{equation}
Generally, the $\chi^{(2)}$ and $\chi^{(3)}$ susceptibility are
at the order of $10^{-12}m/V$ and $10^{-21}m^{2}/V^{2}$, respectively.
Thus the value of $\kappa_{d}\simeq g_{2}^{2}/g_{3}$ should be of
the magnitude of $10^{9}\sim10^{10}\,\mathrm{Hz}$, corresponding
to cavity quality factor $Q$ of the magnitude $10^{4}\sim10^{5}$
at telecommunication wavelength. This condition is realistic for many
integrated photonic platform, such as Aluminum Nitride (AlN) \cite{Guo2016,Guo2016a},
GaAs \cite{gayral1999high}, GaN \cite{bruch2015broadband} and
Lithium Niobate (LN) \cite{wang2014integrated,lin2015fabrication}.
Note that the condition can be relaxed by changing the detuning of
mode $d$ or engineering the mode overlap factors. In Fig.$\:$1(c),
we plot the scaling relation between the coupling strengths of the
cascaded $\chi^{(2)}$-$\chi^{(2)}$ process and $\chi^{(3)}$ process.
The black line shows the situation that the coupling strength $g_{2}^{2}/\kappa_{d}$
of the cascaded $\chi^{(2)}-\chi^{(2)}$ process equals $g_{3}$.
From this figure, it can be seen that the coupling strength $g_{3}$
of the third-order process possibly becomes comparable to or even
higher than the cascaded second-order process, which depends on the
detuning and quality factor of mode $d$. When $g_{3}>g_{2}^{2}/\kappa_{d}$
(i.e., the upper area of Fig.$\:$1(c)), the cascaded $\chi^{(2)}$-$\chi^{(2)}$
process is dominant in the system. 

\section{Model}

In this section, we study the simplified model that only includes the nearest modes $b,c$, to
simplify the complex nonlinear system and reveal the essential physics
of the multimode and multiple nonlinear processes. In presence of
continuous-wave pumping field with frequency near mode $a$, the Hamiltonian
can be written as
\begin{eqnarray}
H & = & \omega_{a,0}a^{\dagger}a+\omega_{b,0}b^{\dagger}b+\omega_{c,0}c^{\dagger}c+\omega_{d,0}d^{\dagger}d\nonumber \\
&  & +g_{3}(a^{\dagger^{2}}bc+a^{2}b^{\dagger}c^{\dagger})+g_{22}(a^{\dagger^{2}}d+a^{2}d^{\dagger})\\
&  & +g_{21}(bcd^{\dagger}+b^{\dagger}c^{\dagger}d)+\epsilon_{a}(ae^{i\omega_{a}t}+a^{\dagger}e^{-i\omega_{a}t}).\nonumber 
\end{eqnarray}
Here, $\omega_{x,0}$ is the resonant frequency of the mode $x$ with
$x\in\{a,b,c,d\}$, $\omega_{a}$ is the pump frequency near the resonance
of mode $a$, $g_{3}$, $g_{22}$ and $g_{21}$ represent the single
photon coupling strength of FWM, SHG and sum-frequency generation.
$\epsilon_{a}=\sqrt{\frac{2\kappa_{a,1}P_{a}}{\hbar\omega_{a}}}$
is the pump field strength, $\kappa_{a,1}$ is the external coupling
rate between cavity and waveguide and $P_{a}$ is the pump power to
mode $a$. Note that the Hamiltonian does not contain self-phase modulation
(SPM) and cross-phase modulation (XPM) terms, which only shift the
frequencies of the resonant modes in our case. Here, we only care
about the steady state behavior of the system. In the rotating frame,
the Hamiltonian can be simplified to,
\begin{eqnarray}
H & = & \delta_{a}a^{\dagger}a+\delta_{b}b^{\dagger}b+\delta_{c}c^{\dagger}c+\delta_{d}d^{\dagger}d\nonumber \\
&  & +g_{3}(a^{\dagger^{2}}bc+a^{2}b^{\dagger}c^{\dagger})+g_{21}(d^{\dagger}bc+db^{\dagger}c^{\dagger})\label{eq:Hamiltonian}\\
&  & +g_{22}(a^{\dagger2}d+a^{2}d^{\dagger})+\epsilon_{a}(a+a^{\dagger})\nonumber 
\end{eqnarray}
where $\delta_{x}=\omega_{x,0}-\omega_{a}$ with $x\in\{a,b,c\}$
and $\delta_{d}=\omega_{d,0}-2\omega_{a}$.

For pump power below the OPO threshold $P<P_{c}$, $b=c=0$ (neglecting
the vacuum induced SPDC and FWM), the Hamiltonian reduces to the form
of simple two-mode SHG. Now we consider the pump power $P>P_{c}$,
under which the energy might flow to mode $b$ and $c$ with $b,c\neq0$.
Following the Heisenberg equation and mean field approximation \cite{Guo2016,Guo2016a},
the dynamics of the operator and the corresponding mean field of each
mode both obey the following form, 
\begin{eqnarray}
\frac{d}{dt}b & = & (-i\delta_{b}-\kappa_{b})b-ig_{3}a^{2}c^{\dagger}-ig_{21}c^{\dagger}d\label{eq:a}\\
\frac{d}{dt}a & = & (-i\delta_{a}-\kappa_{a})a-i2g_{3}a^{\dagger}bc-i2g_{22}a^{\dagger}d-i\epsilon_{a}\label{eq:b}\\
\frac{d}{dt}c & = & (-i\delta_{c}-\kappa_{c})c-ig_{3}a^{2}b^{\dagger}-ig_{21}b^{\dagger}d\label{eq:c}\\
\frac{d}{dt}d & = & (-i\delta_{d}-\kappa_{d})d-ig_{22}a^{2}-ig_{21}bc.\label{eq:d}
\end{eqnarray}
Here, we neglect the fluctuations and only concern the mean field
of each mode. The amplitudes of the fields are treated as complex
numbers. At steady state, $\frac{do}{dt}=0$ with $o\in\left\{ a,b,c,d\right\} $,
and we get four coupled nonlinear equations. Multiplying Eq.$\,$(\ref{eq:a})
by the conjugate of Eq.$\,$(\ref{eq:c}), we get the pump independent
parameter
\begin{equation}
|g_{3}a_{s}^{2}+g_{21}d_{s}|^{2}=|(-i\delta_{b}-\kappa_{b})(i\delta_{c}-\kappa_{c})|,\label{eq:g2g3threshold}
\end{equation}
where $a_{s}$ and $d_{s}$ are the amplitude of the fundamental and SH mode at steady state. The parameter depends only on the detunings and decay rates of mode
$b,c$, which is similar to the OPO threshold \cite{yariv19665a2}.
In absence of the $\chi^{(3)}$ process, the maximum intracavity SH
light amplitude fulfills the threshold relation
\begin{equation}
|g_{21}d_{s}|^{2}=|(-i\delta_{b}-\kappa_{b})(i\delta_{c}-\kappa_{c})|.\label{eq:g2threshold}
\end{equation}
Comparing equation \ref{eq:g2g3threshold} with \ref{eq:g2threshold},
it is found that the interference between $\chi^{(2)}$ and $\chi^{(3)}$
process alters the steady state (maximum) value of $|d_{s}|^{2}$
by the factor $g_{3}a_{s}^{2}$. We can imagine that opposite phases
of $g_{3}a_{s}^{2}$ and $g_{21}d_{S}$ will give larger value of
$|d_{s}|$, compared with the pure $\chi^{(2)}$ process. In this
sense, the $\chi^{(3)}$ process increases the SHG power.

Before calculating the SHG efficiency, we first investigate the threshold
of the OPO. The photons in mode $a$ and $d$ can both provide ``gain''
for $b,c$ via nonlinear interactions. If the ``gain'' can compensate
the decay of mode $b,c$, parametric oscillation in these modes appears.
However, the ``gain'' from the two different paths have different
phases, which may result in constructive or destructive interference
of the cascaded $\chi^{(2)}$-$\chi^{(2)}$ and $\chi^{(3)}$ processes.
As a result, the effective \textquotedblleft gain\textquotedblright{}
determines the threshold of the system. For pump power below threshold,
$b=c=0$, the steady state intracavity photon numbers of $a$ and
$d$ are related by SHG
\begin{equation}
d_{s}=\frac{ig_{22}}{-i\delta_{d}-\kappa_{d}}a_{s}^{2}.\label{eq:shg}
\end{equation}
Increasing the pump power to the threshold $P_{c}$, equation \ref{eq:g2g3threshold}
and \ref{eq:shg} hold simultaneously, we get
\begin{eqnarray*}
	|g_{3}+\frac{ig_{21}g_{22}}{-i\delta_{d}-\kappa_{d}}|^{2}|a_{s}|^{4} & = & |(-i\delta_{b}-\kappa_{b})(i\delta_{c}-\kappa_{c})|.
\end{eqnarray*}
We can further derive the intracavity photon number $A_{s}$ in mode
$a$ at the threshold 
\begin{equation}
A_{s}=|a_{s}|^{2}=\frac{\sqrt{|(-i\delta_{b}-\kappa_{b})(i\delta_{c}-\kappa_{c})|}}{|g_{3}+\frac{ig_{21}g_{22}}{-i\delta_{d}-\kappa_{d}}|^{2}}\label{eq:As}
\end{equation}
and the pump threshold $P_{c}$ of OPO
\begin{eqnarray}
P_{c} & = & \frac{\hbar\omega_{a}}{2\kappa_{a,1}}\times\nonumber \\
&  & \left(\frac{4g^{4}}{\delta_{b}^{2}+\kappa_{d}^{2}}A_{s}^{3}+4g^{2}\frac{\kappa_{a}\kappa_{d}-\delta_{a}\delta_{d}}{\delta_{b}^{2}+\kappa_{d}^{2}}A_{s}^{2}+(\delta_{a}^{2}+\kappa_{a}^{2})A_{s}\right).\nonumber \\
\end{eqnarray}

\begin{figure}
	\begin{centering}
		\includegraphics[width=7cm]{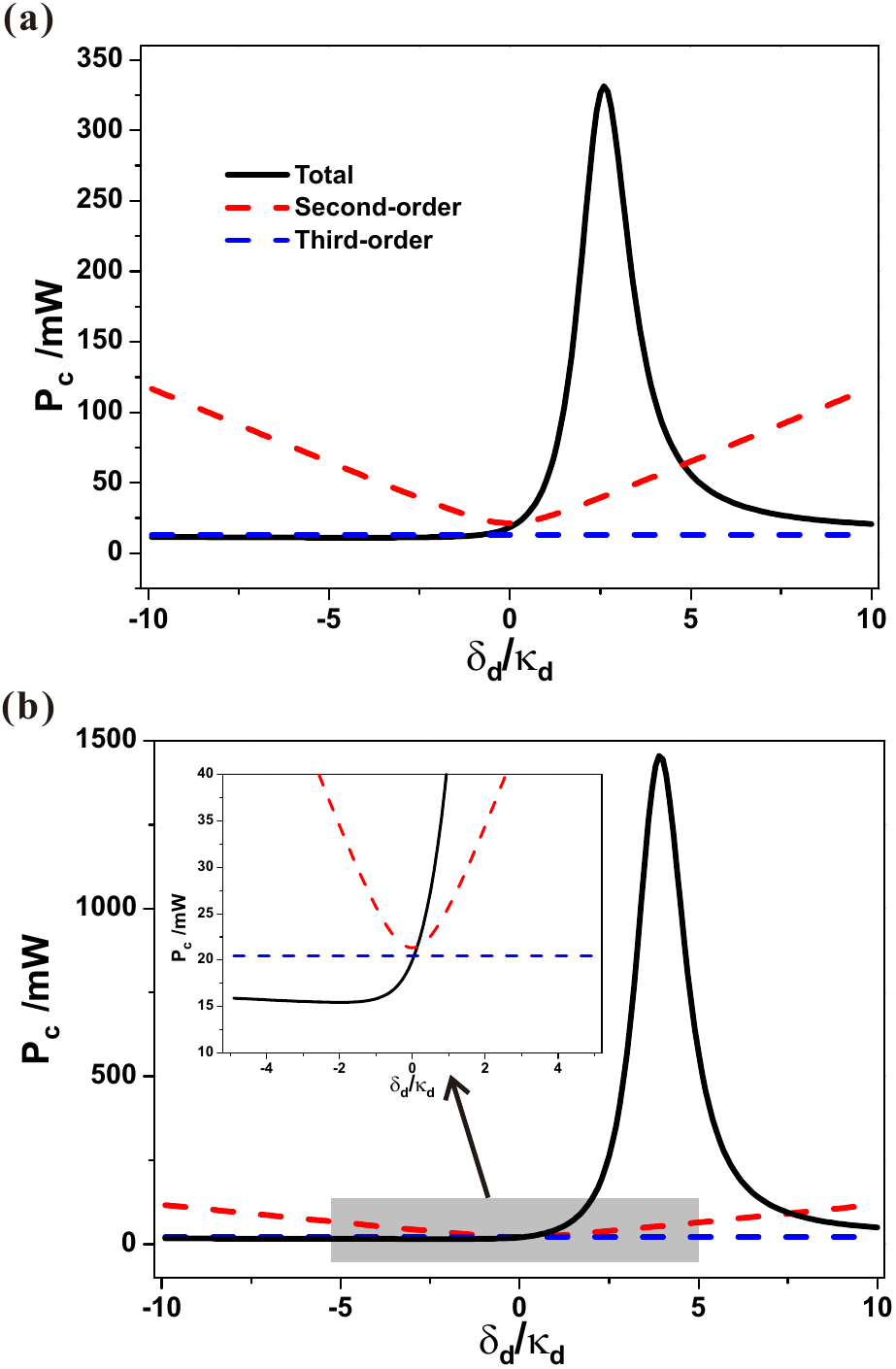}
		\par\end{centering}
	\caption{Relation between the OPO threshold and detuning of the SH mode. (a) The
		parameters are $g_{3}/2\pi=10\:Hz$, $g_{22}/2\pi=0.1\times10^{6}\:Hz$,
		$\kappa_{d}/2\pi=0.8\times10^{9}\:Hz$, which can be realized by the
		current fabrication technology in several platforms. The OPO threshold
		becomes much higher than both the thresholds induced by the $\chi^{(3)}$
		nonlinear process and cascaded $\chi^{(2)}$-$\chi^{(2)}$ process.
		(b) The case for $g_{3}/2\pi=6.5\:Hz$. The inset shows the constructive interference of cascaded $\chi^{(2)}$-$\chi^{(2)}$	and $\chi^{(3)}$ process for $\delta_{d}<0$.
		Dashed Red: OPO threshold of cascaded $\chi^{(2)}$-$\chi^{(2)}$
		process with $g_{3}=0$. Dashed Blue: threshold of $\chi^{(3)}$ process
		with $g_{2}=0$. Black: threshold resulted from interference.}
\end{figure}

Equation (\ref{eq:As}) indicates that the coupling strength $g_{3}$
and cascaded parameter $\frac{ig_{21}g_{22}}{-i\delta_{d}-\kappa_{d}}$
have different phases depending on the second- order coupling strengths
$g_{21}$ and $g_{22}$, the detuning and decay rate of mode $d$.
As shown in Fig.$\:$2(a), the threshold of pump power is altered
by the detuning of the SH mode. For blue detuning of $d$ ($\delta_{d}>0$)
and $g_{3}>0$, $g_{3}$ and $\frac{ig_{21}g_{22}}{-i\delta_{d}-\kappa_{d}}$
have opposite phases, resulting in destructive interference between
the two processes. The destructive interference reduces the effective
``gain'' parameter, thus raising the threshold rapidly. For $\delta_{d}=2\kappa_{d}$,
the threshold increases by $26$ times compared to the case $g_{3}=0$,
and $16$ times compared to OPO induced by only $\chi^{(3)}$process.
While for red detuning, the two paths interfere constructively and
can reduce the threshold (Fig.$\:$2(a)), which may find application in building low-threshold
parametric oscillators and frequency combs.

As a consequence, the generated SH power is changed in presence of
$\chi^{(3)}$ process in two ways. First, for the case $g_{3}=0$,
the SH mode photon number has a upper bound $\frac{|(-i\delta_{b}-\kappa_{b})(i\delta_{c}-\kappa_{c})|}{g_{21}^{2}}$
restricted by OPO threshold \cite{Marte1994}. The change of the
threshold promotes or reduces the upper bound. Second, even though
Eq.$\,$(\ref{eq:g2g3threshold}) holds regardless of the pump power,
the detuning $\delta_{d}$ or the nonlinear coupling among the four
modes may change the phase difference $\Delta\theta$ between $g_{3}a_{s}^{2}$
and $g_{21}d_{s}$ to fulfill $\pi/2<\Delta\theta<3\pi/2$. In this
case, the upper bound from OPO threshold can be broken if the pump
is strong enough.

\begin{figure}
	\centering{}\includegraphics[width=7cm]{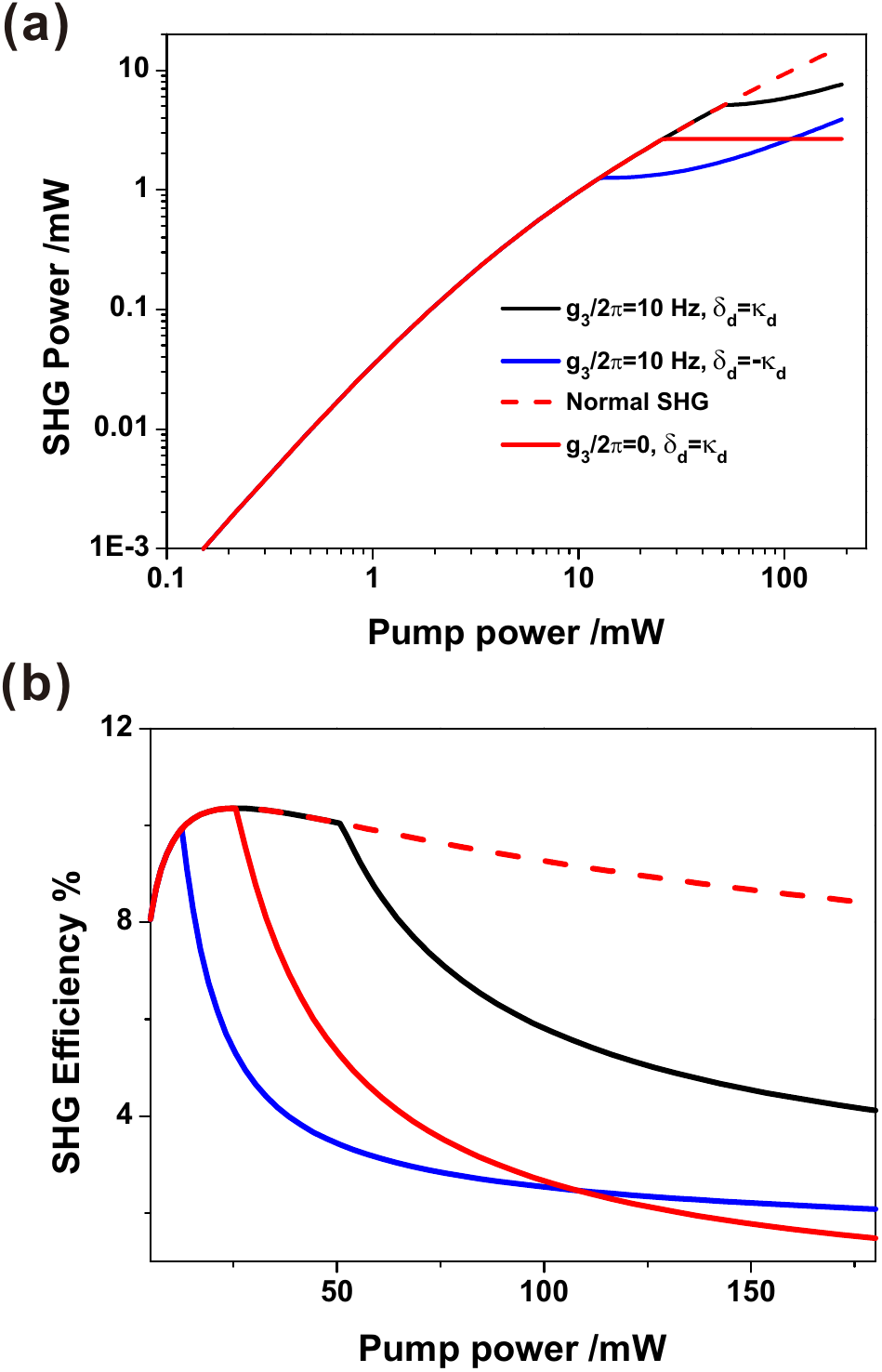}\caption{Relation between the SH mode power (a), SHG efficiency (b) and the
		pump power. Dashed red: Isolated SHG; Solid Red: SHG with $g_{3}=0$.
		Black and Blue: $g_{3}/2\pi=10\:Hz$, and $\delta_{d}=\pm\kappa_{d}$. Normal SHG: SHG only considering the coupling between the fundamental and SH mode.}
\end{figure}

The relationship between the pump power and the output SH power is
obtained by numerically solving the coupled nonlinear equations (Eqs.$\,$(\ref{eq:a})-(\ref{eq:d}))
As shown in Fig.$\,$3, one sees that the threshold changes for different
detunings, which is consistent with the results in Fig.$\:$2. When
the pump power increases beyond the threshold, part of the pump energy
is converted to the neighbor modes of $a$ and the SH power decreases.
When destructive interference happens, the efficiency of the SHG can
evolve along the isolated SHG curve to achieve a higher power,
avoiding the energy flow to other modes in a large range of pump power.
When constructive interference happens, the threshold is even lower
than that of $g_{3}=0$, which accelerates the loss of pump energy.
Another striking phenomenon we should pay attention is the conversion
efficiency for pump powers above the threshold. The intra-cavity photon
number of the SH mode is not restricted by the upper bound in general
OPO process. Both the curves of destructive and constructive interference
(Blue line in Fig.$\:$3) can exceed the upper bound restricted by
the general OPO(Solid red line in Fig.$\:$3). Consequently, the SHG efficiency
can be unambiguously increased by the interference effect. This phenomenon
indicates that the phase relation between $g_{3}a_{s}^{2}$ and $g_{21}d_{s}$
can indeed change along with the increase of the pump, as well as
the change of $\delta_{d}$. The photon number amplitude $d_{s}$
in mode $d$ must increase to compensate the increasing value of $a_{s}^{2}$
to fulfill the conservation condition in equation \ref{eq:g2g3threshold}.
To summarize briefly, the interference can affect the SHG efficiency
by changing the threshold of the system and breaking the restriction
of OPO.

Then, we go further to investigate how to acquire higher SH power
for any given pump power and microcavity. An effective way is to control
the interference of different nonlinear processes to avoid or reduce
the energy flow to the fundamental frequency modes near $a$. Changing
the detuning of mode $d$ to values having different signs with $g_{3}$
can generate the desired destructive interference, which can be realized
using thermal or electro-optic methods. Fig.$\:$4 shows the intracavity
photon number of the SH mode as the frequency detuning $\delta_{d}$
varies. It can be seen that there is a strong correlation between
the signs of $g_{3}$ and $\delta_{d}$. 

\begin{figure}
	\includegraphics[width=7cm]{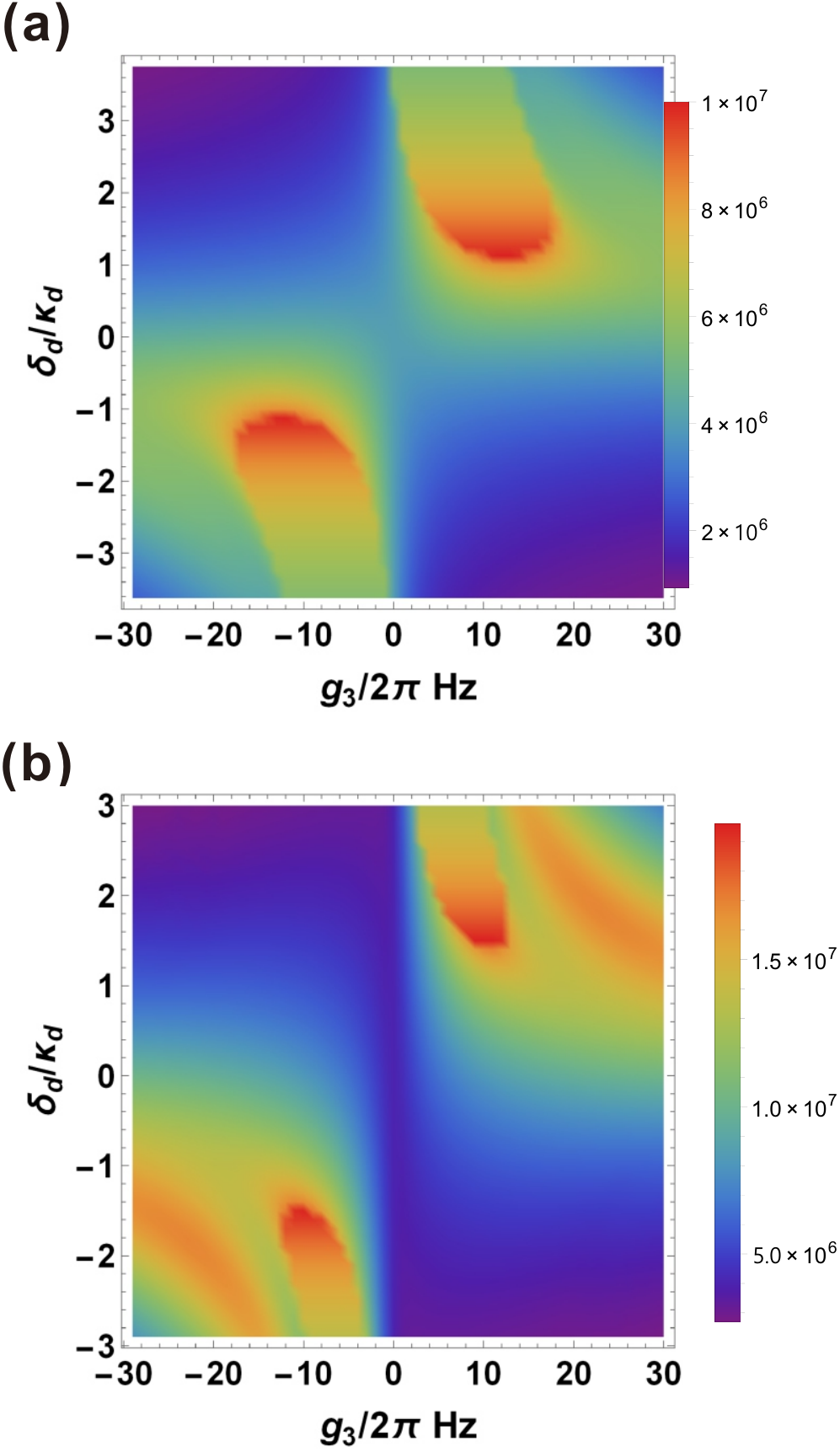}
	
	\caption{Dependences of the intra-cavity photon number in the SH mode on the
		detuning $\delta_{d}$ and the coupling strength $g_{3}$ for different
		pump powers, with $P=70\:mW$ in (a) and $P=200\:mW$ in (b).}
\end{figure}

Since the detuning $\delta_{d}$ between the SH mode and the frequency
of pump laser can modify the phase relation of the two paths, it should
be carefully controlled to improve the conversion efficiency. Note
that zero detuning does not maximize the highest conversion efficiency.
The comparison between Figs.$\,$4(a) and 4(b) indicates that the
detuning for highest conversion efficiency is also related to the
pump power, as well as $g_{3}$. This result demonstrates our deduction
that the phase contrast of the two paths is intensity-dependent.

\section{discussion}

Throughout our simulations, the second-order nonliner coupling strength is $g_{2}/2\Pi=0.1\:Hz$ and the third-order nonlinear coupling strength is at the order of $1-10\:Hz$. The decay rate of the SH mode is $0.8\times 10^{9}\:Hz$, corresponding to loaded quality factor of  $2.5\times 10^{5}$ at $775\:nm$ wavelength. In Ref. \cite{Guo2016,Guo2016a,xiangguo}, the authors reported AlN microrings with diameter of several tens of micrometer and measured the quality factor to be $2.3\times 10^{5}$ and $1.16\times 10^{5}$ in $775\:nm$ and $1550\:nm$ band. The estimated nonlinear coupling strengths are $0.11\:MHz$ for second-order process and several $Hz$ for third-order process.  Since our simulation are based on the current fabrication technology, we believe our prediction could be experimentally realized in the near future.

In the above analysis, we have made several simplifications to the
multimode microcavity with multiple nonlinear processes. The following
processes are neglected: the FWM among the visible modes, SPM, XPM,
sum/difference-frequency generation between the fundamental and SH
modes. These simplifications are valid when the frequencies of the
resonant modes are not equally spaced and the frequency shifts of
the resonant modes due to SPM and XPM are small. For higher pump powers
or microcavities with lower dispersion coefficients, more modes should
be included to make a complete model. In addition, other nonlinear
effects, such as Brillouin scattering, Raman scattering and nonlinear
losses, may also play an significant role under certain conditions,which
opens a new avenue for studying complex coherent nonlinear phenomenon
in optical microcavities. In addition to the coherent interference
between different nonlinear processes studied here, we think many
other new phenomena will be observed experimentally in the near future.
For example, as a reverse process of SHG, the $\chi^{(2)}$-assisted
SPDC in microcavities can also be influenced by $\chi^{(3)}$ nonlinearity.
Another example is the third-harmonic generation, which can be realized
via either $\chi^{(3)}$ or cascaded $\chi^{(2)}$-$\chi^{(2)}$ nonlinear
process. The coherent interference between different nonlinear processes
provides a powerful method to engineer nonlinear photonics in a microcavity
and develop new applications. In addition, the multiple mode microcavity
with multiple nonlinear processes can also serve as a platform for
fundamental studies on the dynamics and stabilities of complex nonlinear
systems.

\section{Conclusion}

In conclusion, we have investigated the coherent interference between
nonlinear $\chi^{(3)}$ and cascaded $\chi^{(2)}$-$\chi^{(2)}$ processes
in an optical cavity. For a multimode microcavity, the efficient SHG
will accompany with the FWM at pump frequency and non-degeneration
optical parametric oscillation generated by the SH light. We first
demonstratethat the FWM and down-conversion can significantly affect
the efficiency of SHG, due to the saturation and the conversion of
the SH light to other modes. Then, we studied the coherent interplay
between those nonlinear processes, and demonstrate that the interference
can suppress the energy leakage from the SH mode to other modes. According
to our analysis, the phase difference between the cascaded $\chi^{(2)}$-$\chi^{(2)}$
and $\chi^{(3)}$ processes depends strongly on the detuning of the
SH mode, which will significantly influence the threshold of the system.
Above threshold, the upper bound of SHG efficiency for pure-$\chi^{(2)}$
nonlinearity can be broken for appropriate phase difference. Our study
clearly explains the mechanism of SHG affected by other nonlinear
processes and complements the blank in the study of SHG with high-power
pump. It can be generalized to more complex systems containing many
resonances which couple with each other via different nonlinear processes,
including $\chi^{(2)}$, $\chi^{(3)}$ and high-order nonlinearities,
as well as Raman and Brillouin scattering processes.
\begin{acknowledgments}
	This work was supported by National Natural Science Foundation of
	China (NSFC) (61725503, 61431166001, 61505195, 61590932, 11774333),
	National Major Research and Development Program (No. 2016YFB0402502),
	Zhejiang Provincial Natural Science Foundation (Z18F050002), China
	Postdoctoral Science Foundation (NO. 2017M621919).
\end{acknowledgments}

\end{document}